\documentclass[prl,twocolumn,floatfix,showpacs]{revtex4}
\usepackage{graphicx}
\usepackage{amssymb,amsfonts,amsmath}
\usepackage{color}
\usepackage{ulem}

\newcommand{\nc}{\newcommand}
\nc{\rnc}{\renewcommand}
\nc{\nn}{\nonumber}

\nc{\g}{\gamma}
\nc{\om}{\omega}

\rnc{\[}{\left[}
\rnc{\]}{\right]}
\rnc{\(}{\left(}
\rnc{\)}{\right)}

\newcommand{\bra}{\langle}
\newcommand{\ket}{\rangle}
\nc{\vac}{|0\ket}
\nc{\vvac}{\bra0|}

\nc{\cd}{\cdots}
\nc{\sm}[2]{\sum_{#1=1}^{#2}}

\nc{\red}{\textcolor{red}}
\nc{\sred}[1]{\textcolor{red}{\sout{#1}}}

\begin{document}
\title{
Exact relaxation dynamics of 
a localized many-body state in the 1D Bose gas 
}

\author{Jun Sato}
\author{Rina Kanamoto}
\author{Eriko Kaminishi}
\author{Tetsuo Deguchi}
\affiliation{Department of Physics, 
Graduate School of Humanities and Sciences, 
Ochanomizu University, 
2-1-1 Ohtsuka, Bunkyo-ku, Tokyo 112-8610, Japan}

\date{\today}

\begin{abstract}
Through an exact method we numerically solve the time evolution of
the density profile for an initially localized state
in the one-dimensional bosons with repulsive short-range interactions.
We show that a localized state with a density notch 
is constructed by superposing one-hole excitations. 
The initial density profile overlaps the plot of the squared amplitude of a dark soliton in the weak coupling regime. 
We observe the localized state collapsing into a flat profile in equilibrium for a large number of particles such as $N=1000$. 
The relaxation time increases as the coupling constant decreases, 
which suggests the existence of off-diagonal long-range order. 
We show a recurrence phenomenon for a small number of particles such as $N=20$.
\end{abstract} 
\pacs{03.75.Kk,03.75.Lm}
\maketitle
Localized wave packets play a fundamental role in quantum mechanics, 
particularly, in the study of dynamical behavior. 
For a massive particle in a free space 
it is possible to construct a localized state such as Gaussian wave packets 
by superposing single-particle eigenfunctions, i.e., plane waves. 
When it evolves in time according to the Schr\"odinger equation, 
the wave packet collapses due to the uncertainty relations \cite{Sakurai}. 
For interacting many-body systems, on the other hand,
it is in general a formidable task to pursue the unitary time evolution
for a given initial state over a sufficiently long time with satisfactory accuracy \cite{Drummond}. 
It is even elusive how to define and construct
an initially localized state of a physical meaning in a system with 
translation symmetry.

The quantum dynamics of interacting many particles has recently
attracted much interest associated with the question of ``relaxation" and
``thermalization" of isolated systems \cite{Rigol}. 
In order to understand the concept of relaxation 
von Neumann's ergodic theorem 
should be useful \cite{Tumulka,Lebowitz,Tasaki,Monnai}. 
Furthermore, 
relaxation dynamics after a sudden quench 
has been studied for integrable systems \cite{Caux}. 
In the studies 
it is also essential to well define an initial state. 
Recent experiments in cold atoms \cite{Kinoshita} enable us to study the isolated
quantum dynamics of initially localized states because of the high
isolation from the environment and controllability of system parameters. 
Thus it is of importance to
theoretically explore how to describe a localized state
and to track its long-time behavior, in both integrable and nonintegrable systems.

In this Letter, we show that a localized many-body state 
with a density notch is constructed 
by superposing the Bethe ansatz eigenstates of a certain type 
for the 1D interacting Bose gas with repulsive delta-function potentials. 
They are given by a series of one-hole excitations, 
which are called Lieb's type II excitations \cite{Lieb-Liniger}. 
In the weak coupling case, 
the initial density profile is quite similar to 
the graph of the squared amplitude of a dark soliton, 
which is a solution of the 
nonlinear Schr{\"o}dinger equation for the classical scalar field. 
We numerically solve the exact time evolution 
of the expectation value of the density operator. 
Observing the movies of the density profile, 
we find that the system with $N=1000$ shows 
relaxation dynamics as if it were thermodynamically large, 
while that with $N=20$ shows 
finite-size effects such as recurrence phenomena. 
The relaxation time of 
the broken symmetry state resembling a dark soliton
increases in the weak coupling case. 
Thus, the localized wave packets are 
more stable as the constant $c$ becomes smaller, which 
suggests the existence of off-diagonal long-range order. 

Let us consider the Hamiltonian of the 
1D interacting bosons with repulsive 
delta-function potentials, called the Lieb-Liniger (LL) model \cite{Lieb-Liniger}: 
\begin{align}
{\cal H}_{\text{LL}} 
= - \sum_{j=1}^{N} {\frac {\partial^2} {\partial x_j^2}}
+ 2c \sum_{j < k}^{N} \delta(x_j-x_k) . 
\end{align}
Here we assume the periodic boundary conditions 
of the system size $L$ on the wavefunctions. 
Hereafter we consider the repulsive interaction: $ c > 0 $. 
The LL model is 
characterized by a single parameter $\g:=c/n$, 
where $n=N/L$ is the density of the particles. 
We fix the particle density as $n=1$ and vary the coupling constant $c$. 
We employ a system of units with $2m=\hbar =1$. 
The unit of time in our simulation is proportional to $L^{-2}$. 

In terms of the canonical Bose field $\hat{ \psi}(x,t)$, 
the LL model corresponds to 
$
{\cal H}_{\text{NLS}} = 
\int_{0}^{L} dx [ \partial_x \hat{ \psi}^{\dagger} \partial_x \hat{ \psi} + 
c \hat{ \psi}^{\dagger} \hat{ \psi}^{\dagger} \hat{ \psi} \hat{ \psi} ] . 
$
The second quantized Hamiltonian ${\cal H}_{\text{NLS}}$ leads to 
the nonlinear Schr{\"o}dinger equation for the canonical Bose field: 
$
i \partial_t \hat{ \psi} = 
 - \partial^2_x \hat{ \psi} 
+ 2c \hat{ \psi}^{\dagger} \hat{ \psi} \hat{ \psi} \, . 
$ 

In the weak coupling limit ($c \to 0$), 
the one-particle reduced density matrix is well approximated 
by the macroscopic wavefunctions: 
$\rho_1(x, y) \simeq \Psi^{*}(x) \Psi(y)$, where $\Psi(x)$ is the eigenfunction 
corresponding to the maximum eigenvalue of $\rho_1(x, y)$. 
In fact, we can numerically show that the largest eigenvalue of $\rho_1$ is much 
larger than the other eigenvalues for small $c$, 
which suggests the existence of off-diagonal long-range order. 
The density operator $\rho(x)$ is defined by the diagonal elements 
of the one-particle reduced density matrix: $\rho(x)= \rho_1(x, x)$. 

There is a well-known conjecture claiming that, in the weak coupling limit, 
the 1D Bose gas, which is a {\it quantum} integrable system, 
should become a {\it classical} integrable system, which is 
described by the Gross-Pitaevskii equation, i.e., 
the nonlinear Schr{\"o}dinger equation for the classical scalar field. 
It was addressed that the mode of classical dark solitons 
is identified with the type II excitations of the LL model
through the coincidence of their dispersion relations 
in the weak coupling limit \cite{Takayama}. 
However, it is still nontrivial to show how the classical solitons are derived 
from the 1D interacting bosons with $c>0$ even in the weak coupling limit. 
For the attractive case ($c <0$), 
this was studied analytically \cite{Wadati}. 
Through numerical simulation, 
the quantum dynamics of dark solitons in the optical lattice system 
has been investigated for the Bose-Hubbard model \cite{MC}. 
The type II excitations are interpreted as quantum solitons in association
with yrast states \cite{Kanamoto}. 

In the LL model, 
the Bethe ansatz offers an exact eigenstate 
with an exact energy eigenvalue 
for a given set of quasimomenta 
$k_1, k_2, \ldots, k_N$ satisfying 
the Bethe ansatz equations 
for $j=1, 2, \ldots, N$: 
\begin{align}
 k_j L = 2 \pi I_j - 2 \sum_{\ell \ne j}^{N} 
\arctan \left({\frac {k_j - k_{\ell}} c } \right) . 
\label{BAE} 
\end{align}
Here $I_j$'s are integers for odd $N$ and half-odd integers for even $N$. 
We call them the Bethe quantum numbers. 
The total momentum $P$ and energy eigenvalue $E$ are written 
in terms of the quasimomenta as 
$P=\sm{j}{N}k_j=\frac {2 \pi} L \sum_{j=1}^{N} I_j$, $E=\sm{j}{N}k_j^2$. 
If we specify a set of the Bethe quantum numbers 
$I_1<\cdots<I_N$, Bethe ansatz equations 
\eqref{BAE} have 
a unique real solution $k_1 < \cdots < k_N$ \cite{Korepin}. 
In particular, the sequence of the Bethe quantum numbers 
of the ground state is given by 
$I_j=-(N+1)/2+j$ for integers $j$ with $1 \leq j \leq N$. 
The Bethe quantum numbers for low lying excitations are 
systematically derived by putting holes or particles in the 
ground-state sequence. 

\begin{figure}[t]
\includegraphics[width=0.6\columnwidth]{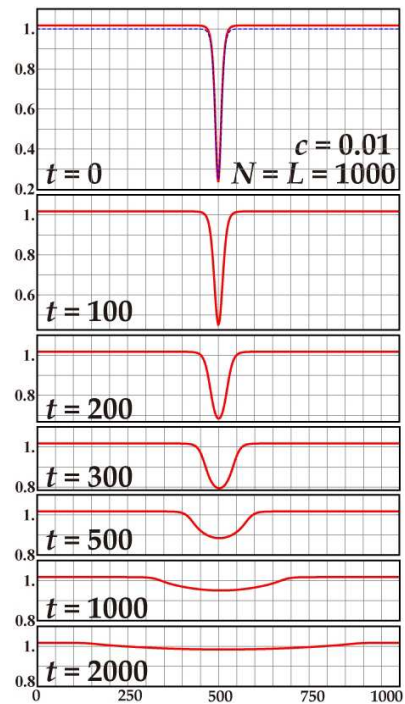}
\caption{(Color online) 
Snapshots of the exact time evolution for the density profile 
with coupling constant $c=0.01$ for $N=1000$, $L=1000$ are shown (red solid lines). 
Also plotted in the top panel is the squared amplitude of a dark soliton (blue dashed line). 
}
\label{snap_shots_c001}
\end{figure}

\begin{figure}[t]
\includegraphics[width=0.6\columnwidth]{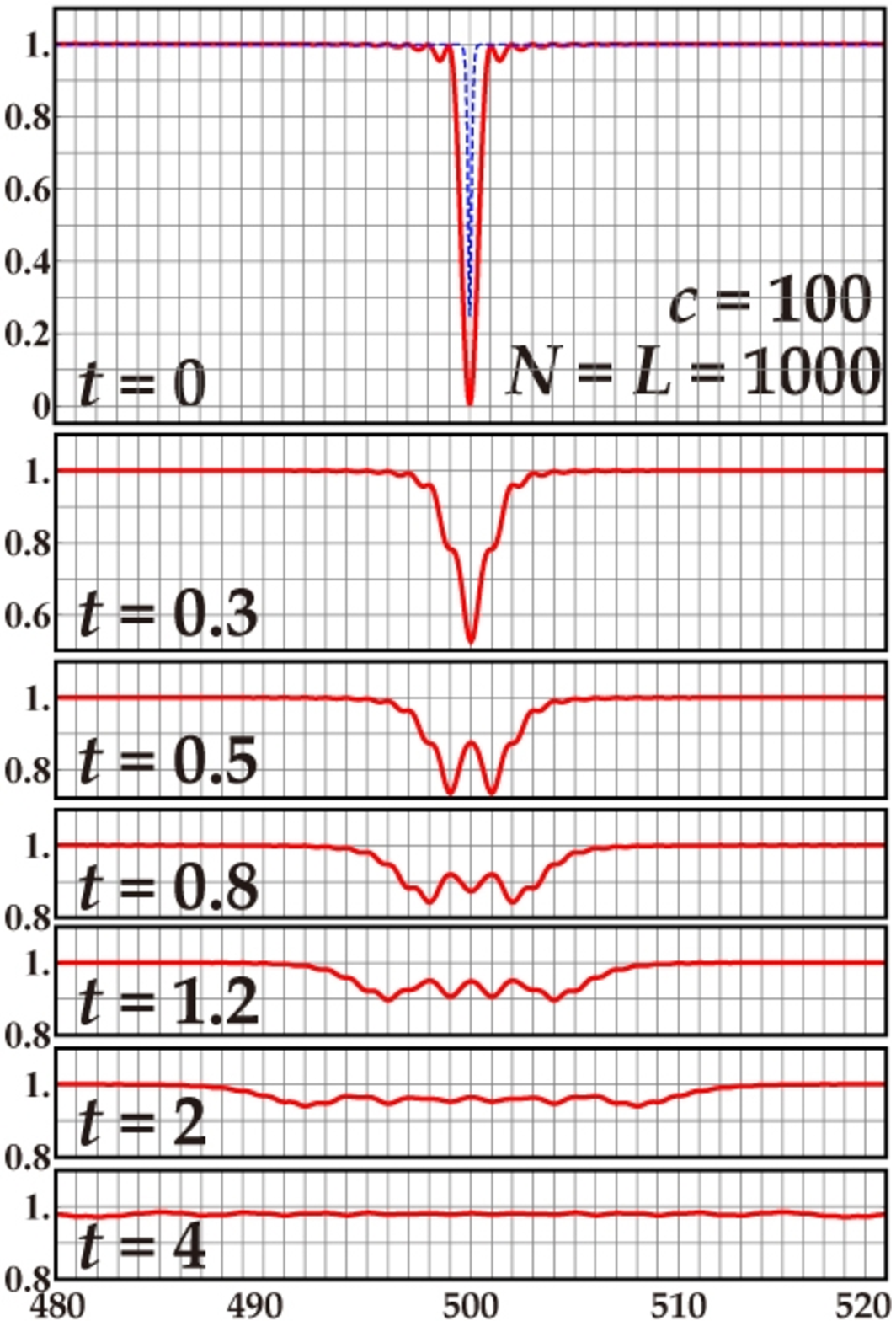}
\caption{(Color online) 
Snapshots of the exact time evolution for the density profile 
with coupling constant $c=100$ for $N=1000$, $L=1000$ are shown (red solid lines). 
Also plotted in the top panel is the squared amplitude of a dark soliton (blue dashed line). 
}
\label{snap_shots_c100}
\end{figure}

\begin{figure}
\includegraphics[width=0.6\columnwidth]{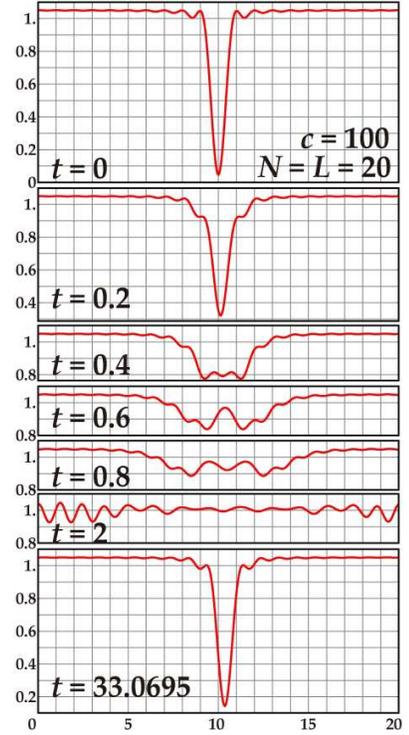}
\caption{(Color online) 
Snapshots of the exact time evolution for the density profile 
with $c=100$ for $N=20$, $L=20$ are shown. 
The solitary dip appears again at $t=33.0695$. 
}
\label{snap_shots_N20}
\end{figure}

Let us now construct an initial state with a localized density profile. 
In the type II branch, we denote by $|P \ket$ 
the normalized Bethe eigenstate of total momentum $P=2 \pi p/L$ 
for each integer $p$ in the set $\{0, 1, \ldots, N-1\}$. 
The Bethe quantum numbers of $|P \ket$ are given by
$I_j=-(N+1)/2+j$ for integers $j$ with $1\leq j \leq N-p$ and 
$I_j=-(N+1)/2+j+1$ for $j$ with $N-p+1\leq j \leq N$. 
For each integer $q$ satisfying $0 \le q \le N-1$, 
we define the coordinate state $|X \ket$ of $X=qL/N$
by the discrete Fourier transformation: 
\begin{align}
| X \ket = \frac 1 {\sqrt{N}} \sum_{p=0}^{N-1} \exp(- 2 \pi i p q/N) \, | P \ket \, . 
\end{align}
It turns out, remarkably, that the density profile of $| X \ket$ 
has a density notch at the position $X+L/2$ 
as shown in the top panels of Figs. 
\ref{snap_shots_c001}, \ref{snap_shots_c100}, and \ref{snap_shots_N20}. 
In the simulation we put $q=0$, and the initial density profile $\rho(x)$ is localized at $x=L/2$.
The ``temperature" of the state $|X(t)\ket$ is estimated by 
equating the mean energy of the state
with the internal energy derived from the thermodynamic Bethe ansatz \cite{Yang}, 
although it is not in thermal equilibrium.

The construction is analogous 
to the analysis of the attractive case ($c<0$) 
in which the Fourier transformation of the $n$-particle bound 
state $|n, P\ket$ with momentum $P$ gives a localized state $|n, X\ket$ 
with the center of mass located at $X$, 
and the matrix element of the field operator 
$\bra n, X'|\hat{ \psi}(x,t)|n+1, X\ket$ 
exactly corresponds to a bright soliton solution 
of the nonlinear Schr{\"o}dinger equation for the classical scalar field with $c<0$ \cite{Wadati}.

Let us introduce $|X(t)\ket = \exp(-i {\cal H}t)|X\ket$. 
We evaluate the expectation value at time $t$ 
of the density operator $\rho(x)=\hat{\psi}^\dagger(x) \hat{ \psi}(x)$
for the initial state $|X\ket$ as 
\begin{align}
\bra X(t)|\rho(x)|&X(t)\ket
=
\frac1N
\sum^{N-1}_{p,p'=0}e^{-2\pi i(p-p')q/N} \nn\\&\times
e^{i(P-P')x-i(E_p-E_{p'})t}
\bra P'|\rho(0)|P\ket, \label{eq:rho(t)}
\end{align}
where $P=2\pi p/L$ and $P'=2\pi p'/L$ denote the total momenta 
of the normalized Bethe eigenstates 
$|P\ket$ and $|P'\ket$, respectively. 

Because of quantum integrability, we numerically obtain 
all the energy eigenvalues $E_p$ of $|P\ket$'s in the type II branch 
and follow the time evolution for quite a long time.

We evaluate the form factor $\bra P'|\rho(0)|P\ket$ in Eq. (\ref{eq:rho(t)}) through 
Slavnov's formula \cite{Slavnov}
together with the
Gaudin-Korepin formula \cite{GK} for the norm of the Bethe eigenstate as 
\begin{align}
&\bra P'|\rho(0)|P\ket
=(-1)^{N(N+1)/2}(P-P')
\(\prod^N_{j,\ell=1}\frac{1}{k'_j-k_\ell}\) \nn\\&\times
\( \prod^N_{j>\ell}k_{j,\ell}k'_{j,\ell}\sqrt{\frac{K(k'_{j,\ell})}{K(k_{j,\ell})} } \)
\frac{\det U(k,k')}{\sqrt{\det G(k)\det G(k')}}, 
\label{eq:Slavnov}
\end{align}
where the quasimomenta $\{k_1,\cdots,k_N\}$ and $\{k'_1,\cdots,k'_N\}$ 
give the eigenstates $|P\ket$ and $|P'\ket$, respectively. 
We use the abbreviations $k_{j,\ell}:=k_j-k_\ell$ and $k'_{j,\ell}:=k'_j-k'_\ell$. 
The kernel $K(k)$ is defined by $K(k)=2c/(k^2+c^2)$. 
The matrix $G(k)$ is the Gaudin matrix, whose $(j,\ell)$ th element is 
$G(k)_{j,\ell}=\delta_{j,\ell}\[L+\sum_{m=1}^NK(k_{j,m})\]-K(k_{j,\ell})$
for $j, \ell=1,2,\cdots,N$. 
The matrix elements of the $(N-1)$ by $(N-1)$ matrix $U(k,k')$ are given by 
\begin{align}
U(k,k')_{j,\ell}&=2\delta_{j\ell}\text{Im}\[\prod^N_{a=1}
\frac{k'_a-k_j + ic}{k_a-k_j + ic}\]
+\frac{\prod^N_{a=1}(k'_a-k_j)}{\prod^N_{a\neq j}(k_a-k_j)} \nn\\&\times
\(K(k_{j,\ell})-K(k_{N,\ell})\). 
\label{eq:matrixU}
\end{align}

We thus numerically solve the exact time evolution of the density profile: 
Once we evaluate the form factors 
at $t=0$ we obtain the density profile at any late time $t$ 
by taking only the sum of exponentials \cite{note}. 
Through Eq. \eqref{eq:Slavnov}, 
the evaluation of the form factors of the Bethe eigenstates with $N$ particles 
is reduced to that of 
the determinants of $N$ by $N$  matrices. 

The density profile of a wave packet for $N=1000$ 
initially has a localized shape without translation symmetry, 
and then it relaxes into a flat profile 
in equilibrium through time development, 
as shown in Figs. \ref{snap_shots_c001} and 
\ref{snap_shots_c100}. 
The relaxation time increases as the constant $c$ decreases. 

In the weak coupling case of $c=0.01$, 
the density profile is smooth and its time evolution is very slow 
as shown in Fig. \ref{snap_shots_c001}. 
The density profile maintains a featureless curve through the time development 
and has a long coherence in space suggesting the existence of off-diagonal long-range order. 

We observe that the initial density profile for $c=0.01$ overlaps 
the plot of the squared amplitude of a dark soliton solution 
of the nonlinear Schr{\"o}dinger equation for the classical scalar field. 
This observation is fundamental for 
making an explicit connection between the type II excitations 
and the dark solitons. 
We shall discuss this in detail in a separate publication. 

In the strong coupling case of $c=100$, 
the time evolution is much faster. 
The initial density profile also shows small oscillating behavior 
at the shoulders of the density notch, 
which are similar to the Friedel oscillations of the Fermi gas. 
The initial wave packet dynamically splits into many fragments, 
which results from the strong repulsive interaction. 
The initial density profile is rather different from 
the graph of the squared amplitude of a dark soliton. 

\begin{figure}
\includegraphics[width=0.6\columnwidth]{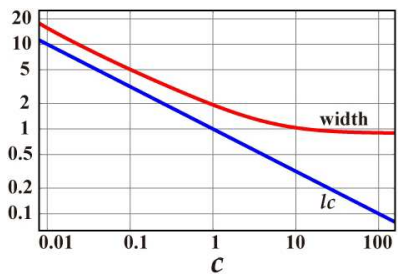}
\caption{(Color online) 
The width of a density notch for the initial density profile
and the healing length $\ell_c=1/\sqrt{cn}$ with $n=1$ 
are plotted against the coupling constant $c$. 
In the evaluation of the width of the density notch, 
we extrapolate the finite $N$ data by $1/N$ expansion and 
estimate the values in the thermodynamic limit $N\to\infty$.
The ratio between width and the healing length approaches a constant value 
in the weak coupling limit $c\to0$. 
}
\label{width}
\end{figure}

The width of the initial density notch 
is approximately proportional to the healing length $\ell_c$ for small $c$, 
as is the case for the classical dark soliton 
(see Fig. \ref{width}). 
For larger $c$, however, it seems that the width 
does not decrease in  proportion to $\ell_c$. 
This suggests the existence of fermionic hard cores. 

Let us calculate the Loschmidt echo, i.e., the fidelity. 
We define it by the overlap between the initial state $|X(0) \ket$ 
and the time-evolved state $|X(t)\ket$ at time $t$ as 
\begin{align}
F(t):=\left|\bra X(t)|X(0)\ket \right|^2=\frac1{N^2}
\left|\sum_{p=0}^{N-1}e^{iE_pt}\right|^2 \, . 
\end{align}

\begin{figure}
\includegraphics[width=0.99\columnwidth]{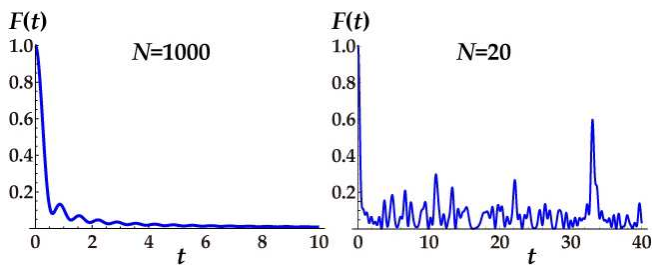}
\caption{(Color online) 
The Loschmidt echo $F(t)=\left|\bra X(t)|X(0)\ket \right|^2$
is plotted against time $t$. 
The number of particles are $N=1000$ (left panel) and $N=20$ (right panel). 
The particle density is $N/L=1$ and
the coupling constant is $c=100$ for both panels. 
 }
\label{fidelity}
\end{figure}

We plot the Loschmidt echo in Fig. \ref{fidelity} 
with fixed particle density $n=N/L=1$ 
in the strong coupling case of $c=100$. 
For a large system size ($N=1000$), 
as shown in the left panel of Fig. \ref{fidelity}, 
it decays almost monotonically 
with the short-time fluctuations being rather small. 
This behavior is common for a wide range of values of $c$ 
with different scales of time. 
In the strong coupling limit $c\to\infty$, 
by sending $N$ and $L$ to $\infty$ with fixed density $n=N/L$ 
we derive analytical forms of the Loschmidt echo: 
$F(t)=[C(n\sqrt{2\pi t})^2+S(n\sqrt{2\pi t})^2]/(2\pi n^2t)$, 
where $C(x)$ and $S(x)$ are given by the Fresnel integrals 
defined by $C(x):=\int_0^x\cos(\pi s^2/2) ds$ and $S(x):=\int_0^x\sin(\pi s^2/2) ds$, respectively. 
As a short-time expansion 
we have $F(t)=1-4\pi^4n^4t^2/45+\mathcal{O}(t^3)$,
and as a long-time expansion 
we have $F(t)=(4\pi n^2t)^{-1}+\mathcal{O}(1/t^2)$. 
The plot of numerical results for $N=1000$ with $c=100$ 
(left panel of Fig. \ref{fidelity}) almost 
overlaps that of the analytical expression of the Fresnel integrals. 
It suggests that the system of $N=1000$ already shows 
thermodynamic behavior. 

For a small system size ($N=20$) the Loschmidt echo is shown in 
the right panel of Fig. \ref{fidelity}. Here, 
we observe large fluctuations due to the finite-size effect.

Let us define a long-time average $\mathbb{E}$ by 
$\mathbb{E}[f(t)]:=\lim_{T\to\infty}\int_0^T dt f(t)/T$. 
Using $\mathbb{E}[e^{i \om t}]=0$ for $\om \neq 0$ and $\mathbb{E}[e^{i \om t}]=1$ for $\om=0$, 
we obtain $\mathbb{E}[F(t)]\sim1/N$ for any value of $c$. 
This is also confirmed by our simulation.

A sharp peak at $t \simeq 33$ in the Loschmidt echo for $N=20$ 
shows the signal of a recurrence phenomenon. 
In fact, the localized wave packet is revived at this time 
after a ``tentative" relaxation as shown in Fig. \ref{snap_shots_N20}. 

In conclusion we have shown that 
the superposition of the Bethe eigenstates 
of the type II excitations leads to a quantum many-body state 
with a localized density profile. 
It perfectly overlaps the plot of the squared amplitude of a dark soliton in the weak coupling regime. 
By means of Slavnov's formula together with the Gaudin-Korepin formula, 
we have numerically solved 
the exact time evolution of the density profile in both the strong and weak coupling cases 
for a large number of particles such as $N=1000$ 
over a very long period of time. 
For a sufficiently large system size ($N=1000$), we observe that 
a density profile 
with broken translation symmetry 
relaxes into a flat profile 
in equilibrium through the time development. 
For a small system size ($N=20$), 
the localized wave packet is revived 
after a tentative relaxation. 
These observations suggest that the method presented in this Letter
should be fundamental for exploring exact approaches to the quantum dynamics of many-body systems. 

The authors thank I. Danshita and K. Sakai for their useful discussions. 
The present research is partially supported by Grant-in-Aid for Scientific Research No. 21710098. 
J.S. is supported by JSPS.


\begin{thebibliography}{99}
\bibitem{note} 
We have solved Bethe ansatz equations \eqref{BAE} 
with errors of $k_j$ given by $10^{-16}$ 
and obtain energy eigenvalues $E_p$ with errors of $10^{-16}$. 
Therefore, the accuracy of the time evolution is assured as long as $t \sim 10^{16}$.
\end{thebibliography}
\end{document}